\documentclass[12pt,a4paper]{article}
\usepackage{jheplikemod,tocloft,amsmath,mathtools,makecell,xcolor}
\let\olditemize\itemize\renewcommand{\itemize}{\vspace{-2pt}\olditemize\setlength{\itemsep}{1pt}\setlength{\parskip}{0pt}\setlength{\parsep}{-0pt}}
\let\oldenumerate\enumerate\renewcommand{\enumerate}{\vspace{-4pt}\oldenumerate\setlength{\itemsep}{1pt}\setlength{\parskip}{0pt}\setlength{\parsep}{0pt}}
\makeatletter\renewcommand\section{\addtocontents{toc}{\protect\addvspace{-15.5\p@}}\@startsection {section}{1}{\z@}{-0.0ex \@plus .2ex \@minus 0.2ex}{1ex \@plus.1ex\@minus 0.5ex}{\normalfont\large\bfseries}}
\renewcommand\subsection{\addtocontents{toc}{\protect\addvspace{-2.25\p@}}\@startsection {subsection}{2}{\z@}{0.5ex \@plus .2ex \@minus 0.2ex}{0.75ex \@plus.1ex\@minus 0.5ex}{\normalfont\bfseries}}
\renewcommand\subsubsection{\addtocontents{toc}{\protect\addvspace{-2.25\p@}}\@startsection {subsubsection}{3}{\z@}{0.5ex \@plus .2ex \@minus 0.2ex}{0.75ex \@plus.1ex\@minus 0.5ex}{\normalfont\bfseries}}
\renewcommand\paragraph{\addtocontents{toc}{\protect\addvspace{-2.25\p@}}\@startsection {paragraph}{4}{\z@}{0.5ex \@plus .2ex \@minus 0.2ex}{0.75ex \@plus.1ex\@minus 0.5ex}{\normalfont\bfseries}}
\definecolor{hred}{rgb}{0.575,0.0,0.225}
\definecolor{rindou1}{rgb}{0.4431,0.2862,0.7960}
\definecolor{rindou2}{rgb}{0.0078,0.1215,0.4392}
\definecolor{lapis}{rgb}{0.0.0470,0.2941,0.5568}
\definecolor{emerald}{rgb}{0.31, 0.78, 0.47}
\definecolor{pinegreen}{rgb}{0.0, 0.47, 0.44}
\definecolor{jade}{rgb}{0.0, 0.66, 0.42}
\definecolor{hblue}{rgb}{0,0,0.575}
\definecolor{hred}{rgb}{0.575,0.0,0.225}
\definecolor{hgreen}{rgb}{0.0,0.4,0.2}
\definecolor{hteal}{rgb}{0.0,0.545,0.7451}

\newcommand{\eq}[1]{\vspace{-4.5pt}\begin{equation}#1\vspace{-4.5pt}\end{equation}}

\newcommand{\fwbox}[2]{\text{\makebox[#1][c]{$\hspace{-150pt}\displaystyle#2\hspace{-150pt}$}}}
\newcommand{\fwboxL}[2]{\text{\makebox[#1][l]{$#2$}}}
\newcommand{\fwboxR}[2]{\text{\makebox[#1][r]{$#2$}}}
\newcommand{\equivR}{\fwbox{14.5pt}{\hspace{-0pt}\fwboxR{0pt}{\raisebox{0.47pt}{\hspace{1.25pt}:\hspace{-4pt}}}=\fwboxL{0pt}{}}}
\newcommand{\equivL}{\fwbox{14.5pt}{\fwboxR{0pt}{}=\fwboxL{0pt}{\raisebox{0.47pt}{\hspace{-4pt}:\hspace{1.25pt}}}}}

\renewcommand{\phi}{\varphi}
\renewcommand{\bar}{\overline}
\renewcommand{\hat}{\widehat}
\renewcommand{\tilde}{\widetilde}

\title{\huge Covariant BCFW for All Masses}
\author[\hspace{-2pt}\dagger,\ast]{\vspace{-24pt} Jacob~L.~Bourjaily\hspace{-1pt},}\emailAdd{jbourj@ias.edu}
\author[\hspace{-4pt}\ast]{\vspace{-2pt} Michael~Plesser~}\emailAdd{plesser@psu.edu}
\affiliation[\dagger]{School of Natural Sciences, Institute for Advanced Study, Princeton, NJ, 08540, USA} 
\affiliation[\ast]{Institute for Gravitation and the Cosmos, Department of Physics,\\Pennsylvania State University, University Park, PA 16802, USA}
\abstract{%
We use the massive spinor-helicity formalism to describe a novel construction for massive spinors which allows one to connect the many choices of bases useful in the representation of amplitudes involving massive particles, and we use this to derive a new formulation for the general BCFW shift that is valid regardless whether or which of the particles involved are massive or massless. This formalism should help facilitate a more coherent implementation of on-shell recursion for computing amplitudes involving massive particles in four dimensions. 
}
\preprint{}
\begin{document}
\maketitle
\pagenumbering{roman}
\setcounter{page}{1}\vspace{0pt}\setcounter{section}{0}
\pagenumbering{arabic}
\section{Introduction}
\label{sec:introduction}

In perturbative Quantum Field Theory, one first learns to calculate scattering amplitudes using the Feynman expansion derived from the theory's Lagrangian. These quantities are needed in a wide array of physical contexts from collider physics at the LHC \cite{Giele:1993dj, Mangano:1990by, Bern:2007dw} to gravitational waveform modeling for LIGO \cite{buonanno2022snowmasswhitepapergravitational, Bern:2019crd, Adamo_2024}. However, the textbook approach is often both morally and practically unsatisfactory: Feynman diagrams are not gauge-invariant, contain spurious singularities, involve the exchange of non-physical off-shell particles, and their number grows explosively with particle number and perturbative order. Nevertheless, one often finds that the final answer is far simpler than would be expected given the complexity of the intermediate stages: when the dust settles, conspiratorial simplifications yield a remarkably simple answer.

There is then understandably great interest in developing methods that manifest this simplicity, and not unreasonable optimism that such methods actually do exist. Indeed, a tremendous amount of fruitful work has already been done in this direction (for starters, see \cite{vanderWaerden:1929, Kleiss:1985yh, Parke:1986gb, Berends:1987me, Duca1999NewCD, Witten:2003nn, BCF, BCFW, Bourjaily:2023uln, Bern:2007dw, Bern:2011qt, Bourjaily:2017wjl, PositiveGrassmannianBook, ArkaniHamed:2010gh, ArkaniHamed:2010kv, Arkani-Hamed:2014dca, Arkani-Hamed:2013jha, BCJ, Bern:2010ue, Bourjaily:2023ycy}). Massless spinor-helicity formalism \cite{vanderWaerden:1929, Kleiss:1985yh} and BCFW-recursion \cite{BCF, BCFW} are two of the most well-known and powerful such tools, emphasizing the little group covariance and pole-structure of amplitudes respectively. They have allowed us to calculate massless amplitudes to orders which would be unthinkable using the traditional approach \`a la Feynman. The virtuous cycle of new development, new data, new insights, new development has been \emph{enormously} fruitful. 

Much of this progress has been made in the context of four-dimensional massless gauge theories, but often one wants to calculate such quantities for \emph{massive} theories, such as QED. Even in the context of massless gauge theory, massive amplitudes can be useful as input to unitarity constructions of loop amplitudes \cite{Bern:2011qt, Bourjaily:2017wjl}. Famously, this gives access to rational terms in pure (non-supersymmetric) Yang-Mills theory which are otherwise difficult to determine \cite{Bern:1995db, Bern:2010qa, Elvang:2011ub}. Unfortunately, the tools relevant for computing massive amplitudes are less mature. A covariant extension of the massless spinor-helicity formalism for massive states was proposed by Arkani-Hamed, Huang, and Huang in \cite{Arkani-Hamed:2017jhn} and BCFW-shifts of massive particles were discussed in \cite{Badger_2005, Schwinn_2007,  Ballav_2021, Wu_2022}, albeit with limitations that we will resolve. 

In this work, we present a new formula for the BCFW-shift in the covariant massive spinor-helicity formalism in four dimensions. The covariant shift generalizes the existing formula for the shift of two massive particles---which relies upon adopting a not-particularly-simple basis for the spinors---to a covariant form valid for any basis of spinors. Furthermore, our formula \emph{reproduces all known BCFW-shifts for any pair of masses as special cases}. We call this the covariant BCFW-shift, and it is the most general answer to the question; ``how can one deform two spinors in a way that preserves the on-shell and momentum conservation conditions, and will yield linear recursion relations?'' Our formula provides a method of shifting massive spinors prepared in any basis, and reveals a basis choice yielding an improved version of the mixed-mass BCFW-shift. Remarkably, we find that---in appropriate bases of spinors---all combinations of masses are shifted in exactly the same way. BCFW-recursion is a machine that can be staggeringly efficient when finely tuned, and covariant methods allow one to leverage the freedom in basis choice for massive spinors to streamline the recursion. To that end, we also present new formulae that trivialize the construction of massive spinors in the covariant framework.
\vspace{\baselineskip}

In section \ref{sec:massive-spinor-review} we review spinor-helicity formalisms and present our new method of constructing massive spinors given a momentum. In section \ref{sec:covariant-bcfw} we discuss BCFW-recursion and present our formula for the covariant shift and its supersymmetrization, demonstrating how it reproduces all known BCFW-shifts as special cases and immediately suggests a new-and-improved form of the shift involving one massive and one massless state. In section \ref{sec:examplitudes} we use the covariant BCFW-shift to compute the same amplitude by shifting different combinations of massive and massless states yielding superficially distinct expressions that are all nevertheless equivalent. We then sketch how this calculation could have been done supersymmetrically and the advantages of doing so. In section \ref{sec:conclusion} we conclude and discuss applications and future directions.\\

\section{Spinor-Helicity Formalism in Four Dimensions}
\label{sec:massive-spinor-review}

First let's motivate why one should even care about spinor-helicity to begin with. In the massless case there is the uniqueness of the helicity basis to be contrasted with the non-uniqueness of an approach based on polarization vectors, but in the massive case there is non-uniqueness on both sides. There is, however, a difference between non-uniqueness and redundancy: to embed massive states in Lorentz tensor fields requires conditions be imposed (\emph{e.g.} tracelessness, antisymmetry, etc) in order to match the correct $2s+1$ physical degrees of freedom. Imposing these conditions is the source of gauge-redundancy, to the perennial bane of students and researchers. In massive spinor-helicity it is true that there is no preferred basis, but we always have exactly the right number of degrees of freedom by construction, and gauge-invariance is guaranteed automatically. Furthermore, we are often interested in massive spin-$\frac{1}{2}$ fermions, and massive spinors are especially apropos in this case. Higher spin massive particles can also be considered by building rank-$2s$ symmetric tensors out of the massive spinors with exactly the right $2s+1$ degrees of freedom \cite{Arkani-Hamed:2017jhn}. 
\vspace{\baselineskip}

\subsection{Review of Massless and Massive Spinor-Helicity}
\label{subsec:review-of-MSH}
The starting point for the massive formalism is the massless one, which we review. For more complete reviews (beware differences in conventions), see (\emph{e.g.}) \cite{vanderWaerden:1929, Elvang:2013cua, Arkani-Hamed:2017jhn}. The short version of the story is that for particle `$i$' with mass $m_i$ and momentum $p_i$ we have a bi-spinorial representation of its momentum, $p_i^{a\dot{a}}\equivR p_i^\mu (\sigma^{a\dot{a}})_{\mu}$, which we think of as a $2\times 2$ matrix for which $\det(p_i) = m_i^2$. A particle for which this is true is called on-shell, and the external particles being scattered are always on-shell.

In Lorentzian signature with $\eta^{\mu\nu} = \text{diag}(1,-1,-1,-1)$ one has
\eq{
(\sigma^{a\dot{a}})_{\mu} = (\begin{pmatrix}1&0\\0&1\end{pmatrix},\begin{pmatrix}0&1\\1&0\end{pmatrix},\begin{pmatrix}0&-i\\i&0\end{pmatrix},\begin{pmatrix}1&0\\0&-1\end{pmatrix}).
}
Often it is convenient to work with $\eta^{\mu\nu} = \text{diag}(1,-1,1,-1)$, in which case we have
\eq{
(\sigma^{a\dot{a}})_{\mu} = (\begin{pmatrix}1&0\\0&1\end{pmatrix},\begin{pmatrix}0&1\\1&0\end{pmatrix},\begin{pmatrix}0&1\\-1&0\end{pmatrix},\begin{pmatrix}1&0\\0&-1\end{pmatrix}).
}
This has the pleasant and convenient effect of rendering $p^{a\dot{a}}$ real-valued for real $p^\mu$.
\vspace{\baselineskip}

A null (rank-1) matrix can be represented as $p^{a\dot{a}} = \lambda^a \otimes\tilde{\lambda}^{\dot{a}} \equivL {}^a|i \rangle [\tilde{i}|^{\dot{a}}$, which are precisely our 2-component Weyl spinors.\footnote{The tilde is not typically present in massless spinor-helicity notation, but will prove helpful in a moment when we pass to the massive formalism and have more spinors to distinguish.} This decomposition is \emph{nearly} unique, but for the fact that we can rescale the angle spinor by any non-zero constant so long as we rescale the square bracket by the inverse. Under an appropriate rescaling the angle/square spinors in Lorentzian signature are complex conjugates of each other, and this property is only maintained if rescaled by a \emph{phase}, reflecting the expected $U(1)$ massless little group.\footnote{There is a standard caveat about $ISO(2)$ versus $U(1)$ related to ``continuous'' vs discrete spin states. Being concerned with the discrete case, $U(1)$ is the relevant group. See \cite{Weinberg:1995mt} for details.} In the alternating signature where the bi-spinorial momentum is real-valued, the spinors are independent and may be rescaled by any non-zero constant, reflecting the non-compactness of the little group. Such distinctions are ultimately rather unimportant to us, as in BCFW-recursion we complexify the momentum anyway.

The angle/square spinors are associated with minus/plus helicities respectively. Spinor indices are lowered using the $SU(2)$ Levi-Civita tensor $\varepsilon$, and Lorentz invariants are formed as determinants, 
\eq{
\langle ij \rangle \equivR \varepsilon_{ab} \lambda_i^a \lambda_j^b, \hspace{2cm} [\tilde{i}\tilde{j}] \equivR \varepsilon_{\dot{a} \dot{b}} \tilde{\lambda}_i^{\dot{a}}\tilde{\lambda}_j^{\dot{b}}.
}
Throughout this work we use Einstein summation convention on repeated indices. To be completely clear, we mean for the $SU(2)$ Levi-Civita tensor $\varepsilon$,
\eq{
\varepsilon_{ab} = \begin{pmatrix}
    0&1\\-1&0
\end{pmatrix} = \varepsilon_{\dot{a}\dot{b}}.
}
The spinors are Weyl in the sense that,
\eq{
p_i^{\dot{a}a}{}_a|i\rangle = |\tilde{i}]^{\dot{a}}\langle ii\rangle  = 0^{\dot{a}},
\hspace{2cm}
p_i^{a\dot{a}}{}_{\dot{a}}|\tilde{i}] = {}^a|i\rangle[\tilde{i}\tilde{i}] = 0^a .
}
These facts are trivial, the determinant of a matrix with identical rows is always 0.
\vspace{\baselineskip}

For the massive case the bi-spinorial $p^{a\dot{a}}$ is no longer rank-1, but instead rank-2. Thus, it can be written as the sum of a \emph{pair} of bi-spinors. We then proceed by doubling the above construction,
\eq{
\begin{array}{lll}
p_i^{a\dot{a}} &= (\lambda_i^1)^a \otimes ( \tilde{\lambda}_i^1)^{\dot{a}} + (\lambda_i^2)^a \otimes ( \tilde{\lambda}_i^2)^{\dot{a}} &\\
&\equivL \begin{pmatrix} {}^a|i\rangle \\ {}^a|i]\end{pmatrix}^I\begin{pmatrix} [\tilde{i}|^{\dot{a}} & \langle \tilde{i}|^{\dot{a}} \end{pmatrix}_I    &\equivL {}^{Ia}|\mathbf{i}\rangle[\mathbf{\tilde{i}}]_I^{~\dot{a}}\\
&= \varepsilon_{IJ} \begin{pmatrix} {}^a|i\rangle \\ {}^a|i]\end{pmatrix}^I  \begin{pmatrix}  -\langle \tilde{i}|^{\dot{a}} \\ ~~~[\tilde{i}|^{\dot{a}} \end{pmatrix} ^J &\equivL \varepsilon_{IJ} {}^{Ia}|\mathbf{i}\rangle[\mathbf{\tilde{i}}]^{J\dot{a}}
\end{array} .
}
Occasionally we find it useful to suppress some or all indices if confusion is unlikely. We now have four Weyl spinors, $|i\rangle, |i], |\tilde{i}], |\tilde{i}\rangle$, which combine in pairs to form Dirac spinors $|\mathbf{i}\rangle,|\tilde{\mathbf{i}}]$ that carry an $SU(2)$ little group index $I,J\in \{1,2\}$ in addition to the spinor indices $a,\dot{a} \in \{ 1,2 \}$, all of which are lowered by the same $SU(2)$ Levi-Civita tensor (but with different types of indices). Dirac spinors include both left and right handed Weyl spinors, and we maintain the convention that angle spinors are left-handed and square spinors are right-handed. They satisfy the Dirac equation:
\vspace{3pt}
\eq{
p_i^{\dot{a}a}|\mathbf{i}\rangle_{a}^{~I} =\alpha m_i |\tilde{\mathbf{i}}]^{\dot{a}I} \hspace{2cm} p_i^{a\dot{a}} |\tilde{\mathbf{i}}]_{\dot{a}J} = \alpha^{-1} m_i |\mathbf{i}\rangle_{~J}^{a}.
}\vspace{3pt}
\noindent The parameter $\alpha$ reflects the normalization, $\det(p_i) = m_i^2 \implies \det(|\mathbf{i}\rangle)\det(|\tilde{\mathbf{i}}]) = m_i^2$, which still leaves a scale ambiguity of $\det(|\mathbf{i}\rangle) = \alpha m_i$ and $\det(|\tilde{\mathbf{i}}]) = \alpha^{-1}m_i$. We call $\alpha=1$ the canonical normalization of the spinors, although we stress that this is \emph{not} the same as a canonical choice of basis. Unlike the massless case this decomposition is far from being even \emph{semi}-unique. 

Since there is no preferred basis, the little group acts as a change of basis from $\{|i\rangle, |i], |\tilde{i}], |\tilde{i}\rangle\} \to \{|i'\rangle, |i'], |\tilde{i}'], |\tilde{i}'\rangle\}$, where in either case the spinors reconstruct the same momentum. In order to clarify the basis choices available, as well as to solve the practical question of generating massive spinors given a momentum, we introduce the first new result of this work.
\vspace{\baselineskip}

\subsection{Generating Massive Spinors: the A/B Method}
\label{subsec:AB-method}

Since the massive little group acts as a change of basis, we want to span the \emph{Lorentz} spinor indices with generic spinors and then leave free parameters in the \emph{little group} spinor indices which embed the massive spinors in the span of the chosen basis. First, choose basis spinors $|e_1\rangle, |e_2\rangle, |\tilde{e}_1], |\tilde{e}_2]$ with $\langle e_1 e_2 \rangle [\tilde{e}_1\tilde{e}_2] \neq 0$ where at most one spinor satisfies $p_i|e_j\rangle = 0$ or $p_i|\tilde{e}_j]=0$, and arrange `covariant basis spinors': 
\eq{
{}^{Ia}|\mathbf{e}\rangle \equivR \begin{pmatrix}
{}^a|e_1\rangle \\
{}^a|e_2\rangle
\end{pmatrix}, \hspace{2cm} {}_I^{~\dot{a}}|\tilde{\mathbf{e}}] \equivR \begin{pmatrix}
{}^{\dot{a}}|\tilde{e}_1] \\
{}^{\dot{a}}|\tilde{e}_2]
\end{pmatrix}.  
}
The spinors for particle `i' with mass $m_i$ and momentum $p_i$ are (non-uniquely)
\eq{
{}^{Ia}|\mathbf{i}\rangle \equivR  A^{I}_{~J} {}^{Ja}|\mathbf{e}\rangle , \hspace{0.5cm}\mathrm{and}\hspace{0.5cm} {}_I^{~\dot{a}}|\tilde{\mathbf{i}}] \equivR B_{I}^{~J} {}_{J}^{~\dot{a}}|\tilde{\mathbf{e}}], }
\eq{
B_I^{~J} = \frac{\varepsilon . A . \langle \mathbf{e}|p_i|\tilde{\mathbf{e}}].\varepsilon}{\det(A)  \langle e_1 e_2 \rangle [\tilde{e}_1 \tilde{e}_2]}, \hspace{0.5cm}\mathrm{or}\hspace{0.5cm} A^I_{~J} = \frac{\varepsilon . B . [ \tilde{\mathbf{e}}|p_i|\mathbf{e}\rangle.\varepsilon}{\det(B) \langle e_1 e_2 \rangle [\tilde{e}_1 \tilde{e}_2]}. \label{covariant-spinor-ABmat-soln}}\\
\noindent The brackets in the middle are, to be completely explicit, computed as
\eq{
{}^I\langle \mathbf{e}|p_i|\tilde{\mathbf{e}}]_J = \begin{pmatrix}
    \langle e_1|p_i|\tilde{e}_1] & \langle e_1|p_i|\tilde{e}_2] \\
    \langle e_2|p_i|\tilde{e}_1] & \langle e_2|p_i|\tilde{e}_2] 
\end{pmatrix}, \hspace{1cm}
{}_J [ \tilde{\mathbf{e}}|p_i|\mathbf{e}\rangle^I = \begin{pmatrix}
    [ \tilde{e_1}|p_i|e_1\rangle & [ \tilde{e}_1|p_i|e_2\rangle \\
    [ \tilde{e_2}|p_i|e_1\rangle & [ \tilde{e}_2|p_i|e_2\rangle 
\end{pmatrix}.
}

\noindent One may choose either $A$ or $B$ for free, and then Equation \ref{covariant-spinor-ABmat-soln} fixes the other one. If we choose the obvious basis, ${}^{Ia}|\mathbf{e}\rangle = \delta^{Ia}$ and  ${}_I^{~\dot{a}}|\tilde{\mathbf{e}}] = \delta_I^{~\dot{a}}$, these reduce to simply
\eq{
{}^{Ia}|\mathbf{i}\rangle \equivR  A, \hspace{2cm} {}_J^{~\dot{a}}|\tilde{\mathbf{i}}] \equivR B,
}
\eq{
B= -\frac{\varepsilon . A .\varepsilon. p_i}{\det(A)}, \hspace{2cm} A= -\frac{\varepsilon . B .\varepsilon. p_i}{\det(B)} .
}
The freedom to choose either $A$ or $B$ for free means we may choose either $|\mathbf{i}\rangle$ or $|\tilde{\mathbf{i}}]$ \emph{entirely at will} (so long as it is non-null), and this fixes the basis choice. The second matrix is determined entirely by the choice of the first and the choice of covariant basis spinors. One particularly trivial parameterization one may take is simply:
\eq{
{}^{Ia}|\mathbf{i}\rangle = \delta^{Ia}, \hspace{2cm}{}_I^{~\dot{a}}|\tilde{\mathbf{i}}] = \delta_{Ia}p_i^{a\dot{a}}.
}

\noindent If we choose to impose one further condition on $A$ (or $B$), we may normalize as
\eq{\det(A) = \frac{m_i}{\langle e_1 e_2 \rangle}\hspace{0.5cm}\leftrightarrow\hspace{0.5cm}\det(B) = \frac{m_i}{[ \tilde{e}_1 \tilde{e}_2 ]}. \label{canonical-normalization-AB}}
This imposes the canonical normalization on the spinors, 
\eq{
\det(|\mathbf{i}\rangle)=m_i=\det(|\tilde{\mathbf{i}}]).
}
One may also wish to impose hermiticity conditions on the spinors in real Lorentzian kinematics, but for alternating signature this is unnecessary and less general. Furthermore, in the context of BCFW we complexify momentum, breaking hermiticity.\\

\subsection{Some Common Basis Choices}
It is useful to be able to prepare massive spinors in arbitrary bases, but to make contact with known results we must know which bases were used in their derivation.
\subsubsection{The High-Energy Basis}
There is a ``High-Energy basis'' of massive spinors \cite{Arkani-Hamed:2017jhn} wherein for $m\to 0$ we find $|i],|\tilde{i}\rangle \to 0$ and $|i\rangle, |\tilde{i}]$ descend onto the massless variables, but more properly this is a \emph{class} of bases which are distinct for $m\neq 0$ but have the same massless limit (modulo the massless little group). It is often said that from expressions in terms of massive spinor brackets one can recover the massless limit by simply `unbold-ing the spinor,'' but this is dangerous advice that is made subtle by momentum-conservation introducing mass-dependence to the spinors' components \cite{Moynihan_2018, burger2019amplitudesgravityclassicaldiscontinuities, burger2020onshellperspectivesmasslesslimit}. 
\vspace{\baselineskip}

\subsubsection{The Coplanar Basis}
The previously known massive BCFW-shift relies on the `coplanar basis', or `simultaneous light-cone decomposition,' the essence of which is to choose a basis for two massive particles' spinors as below \cite{Pittau_1998, Aguila_2004, OPP}.
\eq{
p_i = k + \alpha q, \hspace{0.5cm} p_j = q + \beta k
}
\eq{
p_i^2 = m_i^2, \hspace{0.5cm} p_j^2 = m_j^2, \hspace{0.5cm} k^2=0=q^2, \hspace{0.5cm} k\cdot q \neq 0
}\vspace{0pt}

\noindent Since $k$ and $q$ are null, most of the spinor brackets between $p_i$ and $p_j$ vanish. Our formula applies to spinors prepared in \emph{any} basis, including the coplanar one for which we recover the known result. While this looks rational at first glance, the vectors' components are forced to contain square roots in order to satisfy the constraints. 
\vspace{\baselineskip}

\section{Covariant BCFW-Recursion}
\label{sec:covariant-bcfw}
The idea of BCFW-recursion is to pick out two legs and shift their momenta by a complex parameter $z$ in a way that preserves the on-shell and momentum conservation conditions. This is done by shifting the spinors such that $\hat{p}_a\equivR p_a + z\,r$ and $\hat{p}_b\equivR p_b{-} z\,r$ for a null $r$ which is orthogonal to both $p_a$ and $p_b$. We can then regard the amplitude as a function of $z$, $A(z)$, and at certain values of $z$ the amplitude may become singular. At tree-level the amplitude has only simple poles corresponding to internal propagators going on-shell, and on which the residues are precisely the product of the factorized sub-amplitudes evaluated on the shifted kinematics. To construct the \emph{unshifted} amplitude $A(0)$ we are interested in, we can consider a contour integral encircling the origin counter-clockwise,
\eq{
A(0) = \oint_\gamma \frac{A(z)}{2\pi i z}.
}
Barring obstructions related to so-called `poles at infinity', one can use the Cauchy residue theorem to invert the contour, picking up contour integrals (residues) around all poles (factorizations) in the complex $z$-plane, which we may write (letting $p_L$ denote the sum of the \emph{unshifted} momenta for a given factorization),
\eq{
A(0) = -\sum_{z^*} \frac{\mathrm{Res}(A(z), z^*)}{z^*} = \sum_{z^*} \frac{A_L(z^*)A_R(z^*)}{p_L^2 - m_I^2}.
}
The factorized sub-amplitudes appearing in the residue can themselves be recursed in the same way until reaching three-point amplitudes which are known. For more detailed reviews of BCFW-recursion and on-shell constructability, see \cite{BCF, BCFW, ArkaniHamed:2008yf, Cohen:2010mi}.

A major disclaimer to our discussion is that `on-shell constructability' is \emph{theory-dependent}, and we certainly do not claim that all theories are BCFW-recursable. A trivial and classic counterexample is $g_3\phi^3{+}g_4\phi^4$ theory: if one tried to recurse the four-point scalar amplitude, the quartic coupling contains independent information that is not accessible via factorization. On-shell constructability aside, there is a purely kinematic question, which is how to shift spinors in a way that preserves the on-shell and momentum conservation conditions and yields linear recursion relations.\vspace{\baselineskip}

Since massive spinors carry a little group index, the shift will be given in terms of a tensor $\Xi^{I}_{~J}$. We propose the following upgrade of the massless BCFW-shift:
\begin{align}
&|\hat{i}\rangle = ~|i\rangle + z|j\rangle, \hspace{2cm} |\hat{\tilde{j}}]=~|\tilde{j}] - z |\tilde{i}], \\
\implies {}^I&|\hat{\mathbf{i}}\rangle = {}^I|\mathbf{i}\rangle + z \Xi^{I}_{~J}{}^J|\mathbf{j}\rangle, \hspace{1cm} {}_J|\hat{\tilde{\mathbf{j}}}] = {}_J|\tilde{\mathbf{j}}] - z \Xi^{I}_{~J}{}_{~I}|\tilde{\mathbf{i}}].
\end{align}
Satisfying momentum conservation is automatic, but the on-shell condition must be solved in terms of spinor brackets, and in order to ensure linear (not quadratic) recursion relations, the tensor must be null: $\det(\Xi)=0$. We have three constraints for four unknowns, which fix up to an overall rescaling:
\begin{equation}
\Xi^1_{~1}=1, \hspace{0.75cm}\Xi^{1}_{~2} = \frac{\langle j | p_i | \tilde{j}\rangle}{\langle j | p_i | \tilde{j}]-\overline{\Delta}},  \hspace{0.75cm} \Xi^{2}_{~1} = \frac{[i|p_j|\tilde{i}]}{ \langle i | p_j | \tilde{i}]-\overline{\Delta}},  \hspace{0.75cm}\Xi^{2}_{~2} = \Xi^{1}_{~2} \Xi^{2}_{~1},
\end{equation}
\vspace{-20pt}
\begin{align}
\overline{\Delta} :&= \Delta - (p_i \cdot p_j),\\
\Delta :&= \sqrt{(p_i \cdot p_j)^2 - m_i^2m_j^2}.
\label{eqn:mbcfw}
\end{align}
\emph{N.b.} $p_i\cdot p_j$ that appears in the above is the \emph{four-vector} dot product, $p_i\cdot p_j\equivR\eta_{\mu\nu}p_i^{\mu}p_j^{\nu}$. The conjugate shift is obtained by simply exchanging $i$ and $j$. The rescaling has been chosen so as to match the usual massless BCFW. For \emph{either} particle being massless, $\overline{\Delta} = 0$, and so this expression is smooth as $m_i, m_j\to 0$. Note that $\Delta^2$ is the Gram determinant of the two momenta. Furthermore, this has a geometric meaning \cite{Damour_2016},
\eq{
\Delta^2 = |p_i \wedge p_j|^2.
}
The appearance of the wedge product is not totally surprising, knowing that we must shift the momenta by a null vector $r$ that is orthogonal to both $p_i$ and $p_j$. 
\vspace{\baselineskip}

\begin{table}[t]
\begin{large}
\begin{center}
\begin{tabular}{|c||c|c|c|c|}
\hline 
\text{Masses} & (0, 0) & ($m_i$, 0) & (0, $m_j$) & ($m_i$, $m_j$)\\
\hline
 $\Xi^I_{~J}$ & $\begin{pmatrix}
1 & 0\\
0 & 0\\
\end{pmatrix}$  & $\begin{pmatrix}
\langle ij \rangle & 0\\
[ ij \rangle & 0\\
\end{pmatrix}$ & $\begin{pmatrix}
[ \tilde{i}\tilde{j} ] & [ \tilde{i} \tilde{j} \rangle\\
0 & 0\\
\end{pmatrix}$  & $\begin{pmatrix}
1 & 0\\
0 & 0\\
\end{pmatrix}$ \\ 
\hline 
Basis & \makecell{High-Energy\\(both)} & \makecell{High-Energy\\(particle $j$)} & \makecell{High-Energy\\(particle $i$)} & \makecell{Coplanar\\(both)} \\
\hline
\end{tabular}
\end{center}
\end{large}
\caption{Summary of known BCFW-shifts and the basis of spinors they use.}
\label{shift-table}
\end{table}

One improvement that can be made in the mixed-mass cases would be to prepare the massive momentum in a basis of spinors so as to make the off-diagonal element of the shift tensor vanish. In other words, we propose using bases for the massive-massless shift where $|i] \propto |j\rangle$ or $|\tilde{j}\rangle\propto |\tilde{i}]$ for the shifts $\langle \mathbf{i} \tilde{j}]$ and $\langle i \tilde{\mathbf{j}}]$ respectively, and it is easy to accomplish such a basis choice using the $A$/$B$ method. We can also rescale the shift tensor to set $\Xi^1_{~1}=1$. Using this proposed basis, the BCFW-shift simplifies, with all other spinors remaining unshifted, to
\eq{
|\hat{i}\rangle = |i\rangle + z |j\rangle, \hspace{2cm}
|\hat{\tilde{j}}]= |\tilde{j}] - z |\tilde{i}].
\label{special-bases-BCFWshift}
}
This is the same form of the shift as for the (usual) massless BCFW-shift, as well as the previously known massive shift in the coplanar basis \cite{Wu_2022}. Table \ref{shift-table} summarizes the known BCFW-shifts and their associated bases.
\vspace{\baselineskip}

It is perhaps useful to define $\Xi^{I}_{~J}\equivL\tilde{\xi}^I \xi_J
$, and then $|\xi . \mathbf{i} \rangle \equivR \xi_J^{~J} |\mathbf{i}\rangle$ and $|\tilde{\xi} . \tilde{\mathbf{j}} ] \equivR \tilde{\xi}^I_{~I} |\tilde{\mathbf{j}} ]$.
Reading off the vectors $\xi$ and $\tilde{\xi}$ is trivial from how we wrote the shift tensor $\Xi$,
\eq{
\tilde{\xi}^I = (1, \Xi^2_{~1}), \hspace{2cm} \xi_J = (1, \Xi^1_{~2}).
}
For example, the covariant BCFW-shift can be written,
\begin{align}
{}^{Ia}&|\hat{\mathbf{i}}\rangle = {}^{Ia}|\mathbf{i}\rangle + z \tilde{\xi}^{I}{}^{~a}|\xi.\mathbf{j}\rangle, \hspace{2cm} {}_J^{~\dot{a}}|\hat{\tilde{\mathbf{j}}}] = {}_J^{~\dot{a}}|\tilde{\mathbf{j}}] - z \xi_J{}^{~\dot{a}}|\tilde{\xi}.\tilde{\mathbf{i}}],
\end{align}
and the shift momentum $r$ is given compactly by, 
\eq{r = (\tilde{\xi}^I\xi_I)|\xi.\mathbf{j}\rangle[\tilde{\xi}.\tilde{\mathbf{i}}|=\Xi^{I}_{~J} {}^{~J}|\mathbf{j}\rangle [\tilde{\mathbf{i}}|_K \Xi^K_{~I}.}
The $z_L$ which sets an internal line on-shell ($p_I = \hat{p}_L$, $p_L = \hat{p}_L(z=0)$, and $p_I^2=m_I^2$) is 
\eq{
z_L = \frac{m_I^2 - p_L^2}{\tilde{\xi}\!\cdot\!\xi \langle \mathbf{j}.\xi|p_L|\tilde{\xi}.\tilde{\mathbf{i}}]}.
}
Notice that the factor $\tilde{\xi}\cdot\xi$ cancels out when one computes the ratio of factors in $z_Lr$. \\

\subsection{Supersymmetric Massive BCFW-recursion}
In supersymmetric theories, \emph{super}amplitudes combine many different component amplitudes into a unified object. For $\mathcal{N}=4$ sYM, supersymmetric BCFW-recursion of the full \emph{super}amplitude is by far the best way to obtain tree-level component amplitudes. Aside from providing many amplitudes from a single recursive calculation, the recursion is \emph{easier} for $\mathcal{N}=4$ sYM because of the absence of poles at infinity, to be contrasted with pure Yang-Mills which does have bad-shifts which generate poles at infinity. Since at tree-level all less-supersymmetric amplitudes (including those of pure Yang-Mills) are identical to the component amplitudes of maximal sYM, the most efficient path to them is via supersymmetric BCFW recursion \cite{Drummond:2008cr, Dixon:2010ik, Bourjaily:2010wh, Bourjaily:2023uln}. \vspace{\baselineskip}

Massive multiplets and their on-shell description in supersymmetry have been discussed in \cite{Herderschee_2019_mos}, and \emph{massive} supersymmetric BCFW-recursion in \cite{Herderschee_2019_cba, Abhishek_2022}. This is essentially using the old, non-covariant massive-massive shift (in the fixed, coplanar basis), so we can upgrade this almost automatically. For supersymmetric recursion, one introduces a quantity called \emph{supermomentum},
\eq{
Q_i \equivR {}^I|\mathbf{i}\rangle \eta_{iI} .
\label{eqn:Q}
}
There is also a tilde version defined similarly, $\tilde{Q}_i \equivR {}_I|\tilde{\mathbf{i}}] \tilde{\eta}_{i}^{~I}$.
The name comes from the obvious analogy with momentum, where we have substituted the square-spinor for the Grassmann-valued $\eta_{iI}$.\footnote{\emph{N.b.} the index $I\in\{1,2\}$, absent in the massless case. The other label $i$ is the particle index.} Superamplitudes conserve supermomentum, and one can ``square'' that relation and this should also vanish, 
\eq{
\delta^{(2)}(Q)\equivR \sum_{i\leq j} {}^I\langle \mathbf{i} \mathbf{j}\rangle^J \eta_{iI} \eta_{jJ}.
\label{eqn:Q2}
}
One can impose supermomentum conservation by making the superamplitude proportional to this fermionic delta, 
\eq{
\mathcal{A} \propto \delta^{(2)}(Q) .
}
We can then investigate how this object is changed under our massive BCFW-shift. Since we want to avoid poles at infinity, we want this object to remain unshifted. Sadly the fermionic delta \textit{is} shifted, $\delta^{(2)}(\hat{Q}) \neq \delta^{(2)}(Q)$, but this can be easily fixed. The trick to preserving supermomentum-conservation is obvious once you remember that we already know how to preserve \emph{momentum}-conservation. We simply complement the shift of the spinors \ref{eqn:mbcfw} with the following shift of the $\eta$'s and $\tilde{\eta}$'s, 
\begin{align}
 \hat{\tilde{\eta}}_{i}^{~I} = \tilde{\eta}_{i}^{~I} + z \Xi^{I}_{~J} \tilde{\eta}_{j}^{~J}, \hspace{2cm} \hat{\eta}_{jJ} = \eta_{jJ} - z \Xi^{I}_{~J} \eta_{iI} .
 \label{eqn:smbcfw}
\end{align}\vspace{\baselineskip}

\section{An Examplitude}
\label{sec:examplitudes}

Before we can recurse we need some three-point amplitudes. The construction of three-point amplitudes for any mass and spin is described in \cite{Arkani-Hamed:2017jhn}, but we will only consider the minimal coupling for two same-mass spin-$s$ particles and one massless helicity $\pm h$ particle. These amplitudes are written in terms of massive spinors as:

\eq{
A[\mathbf{1}^{I_1\cdots I_{2s}},\mathbf{2}^{J_1\cdots J_{2s}},3^{h_3}] = 
\frac{x^{h_3}}{m^{2s-1}} \begin{dcases}
\langle \mathbf{1}\mathbf{2}\rangle^{\odot 2s}, & h_3=+h\\
~[\tilde{\mathbf{1}}\tilde{\mathbf{2}}]^{\odot 2s}, & h_3=-h\\
\end{dcases}.
\label{min-coupling-amps}
}
The $x$-factor (which is non-local and is written using an auxiliary spinor) is given by 
\eq{
x = \frac{\langle q | p_1|\tilde{3}]}{m\langle q 3 \rangle} = -\frac{\langle q | p_2|\tilde{3}]}{m\langle q 3 \rangle}.
}
The choice of spinor $|q\rangle$ is arbitrary, and the $x$-factor is in fact independent of $|q\rangle$. 

As an example, for massive scalars there are no spin-states and we get simply  
\eq{
A[\mathbf{1}^0,\mathbf{2}^0,3^{\pm 1}] = mx^{\pm 1}.
}
For two spin-$1/2$ fermions coupled to a plus-helicity photon (or gluon), we have
\eq{
A[\mathbf{1}^{I},\mathbf{2}^{J},3^{+1}] = x \langle \mathbf{12}\rangle^{IJ} =x\begin{pmatrix}
    \langle 12 \rangle & \langle 12 ] \\
    [ 12 \rangle & [12]
\end{pmatrix}^{IJ}.
}

Finally, let's look at a four-point example. Consider massive adjoint scalars minimally coupled to gluons, where we can compute tree-level colour-ordered partial amplitudes. These can be interpreted as scalar-QCD amplitudes, or component amplitudes on the Coulomb branch of $\mathcal{N}=4$ sYM. Take the amplitude $A[\mathbf{1}^0,2^{+1},3^{+1},\mathbf{4}^0]$. This can be calculated by either a $\langle2 \tilde{3}]$ shift, a $\langle2\tilde{\mathbf{1}}]$ shift, or a $\langle\mathbf{1}\tilde{\mathbf{4}]}$ shift. This presents a puzzle, as the massless and mixed-mass shifts are rational but the massive one is algebraic: how does the square root get resolved if all three calculations are to agree?\vspace{\baselineskip}

Consider first the massive shift $\langle\mathbf{1}\tilde{\mathbf{4}}]$. There is a single factorization to consider,
\eq{
A[\mathbf{1}^02^{+1}3^{+1}\mathbf{4}^0] = \frac{A[\hat{\mathbf{1}}^0,2^{+1},\mathbf{I}^0]A[-\mathbf{I}^0, 3^{+1},\hat{\mathbf{4}}^0]}{p_{12}^2-m^2}.
}
These amplitudes are just multiplications of $x$-factors evaluated on the shifted kinematics and divided by the unshifted propagator. The square-root appears in the shift of $p_1$ and $p_4$ through the $\xi$ and $\tilde{\xi}$ vectors. Both sub-amplitudes contain the square root in multiple places, and only when they are multiplied does the product rationalize. The dependence on the reference spinors $|q_i\rangle$ (which do not have to be the same for both sub-amplitudes), also drops out in a non-trivial way. \vspace{\baselineskip}

Next consider the massless $\langle 2 \tilde{3}]$ shift, which again has a single factorization,
\eq{
A[\mathbf{1}^02^{+1}3^{+1}\mathbf{4}^0] = \frac{A[\mathbf{1}^0,\hat{2}^{+1},\mathbf{I}^0]A[-\mathbf{I}^0, \hat{3}^{+1},\mathbf{4}^0]}{p_{12}^2-m^2}.
}
This is almost identical to the previous result: the sub-amplitudes are again simply $x$-factors evaluated on shifted kinematics, but now everything is manifestly rational (assuming one is using rational kinematics) because of the vanishing of $\Delta$ in $\Xi$.\\

Lastly we can consider the mixed-mass shift $\langle2\tilde{\mathbf{1}}]$,\footnote{\emph{N.b.} that the conjugate shift $\langle \mathbf{1} \tilde{2}]$ yields the \emph{wrong} answer, it is a \emph{bad} shift.} in which case we have the factorization which is instead mediated by a massless gluon exchange,\footnote{One must sum over both helicity-exchanges, but $A[+++]=0$, so we write only one term.}
\eq{
A[\mathbf{1}^02^{+1}3^{+1}\mathbf{4}^0] = \frac{A[\mathbf{4}^0,\hat{\mathbf{1}}^0,I^{+1}]A[-I^{-1}, \hat{2}^{+1},3^{+1}]}{p_{14}^2}.
}
This expression is more substantially different from the other two, but the shift is rational and the right side of the factorization contains a fully massless amplitude, and so the answer is manifestly rational. \vspace{\baselineskip}

These ostensibly distinct expressions were checked to be mutually consistent with each other (a strong check on correctness) and equivalent to the known expression (an incontrovertible check on correctness) \cite{Badger_2005, Ochirov_2018}:
\eq{
A[\mathbf{1}^02^{+1}3^{+1}\mathbf{4}^0] =\frac{ m^2 [\tilde{2}\tilde{3}]}{\langle23\rangle(p_{12}^2-m^2)}.
}

It is remarkable that the $\langle\mathbf{1}\tilde{\mathbf{4}}]$ shift, for all of its square roots and complexity, produces an expression equivalent to such a simple answer. Indeed, we find for $\hat{x}_L\hat{x}_R$ 
\vspace{2pt}
\eq{ 
\frac{\langle q_1| (p_1 + (\frac{m^2-p_{12}^2}{\langle \mathbf{4}.\xi|p_{12}|\tilde{\xi}.\tilde{\mathbf{1}}]})|\xi.\mathbf{4}\rangle[\tilde{\xi}. \tilde{\mathbf{1}}|) |\tilde{2}]}{m\langle q_1 2\rangle}
\frac{\langle q_2| (p_1 + (\frac{m^2-p_{12}^2}{\langle \mathbf{4}.\xi|p_{12}|\tilde{\xi}.\tilde{\mathbf{1}}]})|\xi.\mathbf{4}\rangle[\tilde{\xi}. \tilde{\mathbf{1}}|) |\tilde{3}]}{m\langle q_2 3\rangle} = \frac{[\tilde{2}\tilde{3}]}{\langle 23 \rangle}.
}
\vspace{2pt}

As promised, the left-hand side is independent of the choice of $|q_1\rangle$ and $|q_2\rangle$, the choice of basis for the massive spinors, and is free of any square roots coming from $\xi$ and $\tilde{\xi}$. Calculations with more gluons work out similarly, and the massive versus massless shifts are found to produce equivalent (although not identical\footnote{They could be made identical by judicious application of Schouten and momentum conservation identities, but this is a difficult reduction to automate, and so generally one does not try to directly reconcile such expressions. The easiest way to quickly convince oneself that the expressions agree is via numerical evaluation with rational kinematics.}) expressions. The rationalization of the square root is encouraging, as square roots can cost orders of magnitude in evaluation time. The fact that the square root vanishes from the final answer means one can hope to deal with it symbolically and only substitute in the evaluated expression at the end of the calculation.\vspace{\baselineskip}

\subsection{A \emph{Super}-Examplitude}
Let us sketch the idea of supersymmetric massive recursion. First notice the similarity of the scalar and fermionic three-point amplitudes. Indeed, we can rewrite these as a unified superamplitude for an $\mathcal{N}=1$ massive chiral multiplet $\Phi$ involving a complex scalar $\varphi$ and a chiral fermion $\psi^I$, 
\eq{
\Phi = \phi + \psi^I \eta_I + \frac{1}{2} \phi^* \eta_I\eta^I.
}
For the coupling of the chiral multiplet to the plus-helicity gluon, we find
\eq{
A[\Phi, \Phi, 3^+] = x \delta^{(2)}(Q).
}
To spell out how this encodes the scalar and fermion amplitudes explicitly, expand
\vspace{2pt}
\eq{
\begin{array}{ll}
\delta^{(2)}(Q) &= \sum_{i<j}^2 {}^I\langle\mathbf{i}\mathbf{j}\rangle ^J\eta_{iI}\eta_{jJ} + \sum_{i=1}^2\det(|\mathbf{i}\rangle)\eta_{i1}\eta_{i2} \vspace{5pt} \\ 
 &= \langle 12\rangle \eta_{11}\eta_{21} + \langle 12] \eta_{11}\eta_{22} + [12\rangle \eta_{12}\eta_{21}  + [12] \eta_{12}\eta_{22} \\ 
 &\phantom{=} + \det(|\mathbf{1}\rangle) \eta_{11}\eta_{12} + \det(|\mathbf{2}\rangle) \eta_{21}\eta_{22} .
\end{array}}\vspace{2pt}

To extract the component amplitude for, say, $A[\mathbf{1}^{1/2},\mathbf{2}^{-1/2},3^+]$ we would pick out the component with $\eta_{12}\eta_{21}$, from which we obtain $A[\mathbf{1}^{1/2},\mathbf{2}^{-1/2},3^+] = x [12\rangle$, which is nothing more than $x~{}^{I=2}\langle \mathbf{1} \mathbf{2}\rangle^{J=1}$. Similarly for a scalar component, \emph{e.g.} $A[\mathbf{1}^{0}, \mathbf{2}^{0^*}, 3^+]$ we look at the term involving $\eta_{21}\eta_{22}$, finding
\eq{
A[\mathbf{1}^0,\mathbf{2}^{0^*},3^+] = x \det(|\mathbf{2}\rangle) \propto mx.
}
Had we adopted the canonical normalization we would recover exactly the quoted result from \cite{Arkani-Hamed:2017jhn}, and otherwise it would be different by just an overall constant. If we had taken the ``other'' scalar component, the only difference would have been $\det(|\mathbf{1}\rangle)$ appearing instead, but recall that we've already restricted to $m_1=m_2$. This is one of many instances where factors of `$m$' in cited formulae are \emph{really} basis-dependent spinor brackets which are only equal to $m$ in certain bases. \emph{Caveat lector}: covariant spinors require a degree of care when interpreting formulae found in the literature which may have been derived assuming certain basis choices.\vspace{\baselineskip}

A very clean example can be made in the case of an amplitude with only two massive legs, which are always the ones shifted. This guarantees the factorization sits on the pole associated to the exchange of a massive scalar/fermion. Integrating over the Grassmann variables of the internal exchange shows that,
\eq{
\int d^2 \eta_I \delta^{(2)}(\hat{Q}_L)\delta^{(2)}(\hat{Q}_R) \propto \delta^{(2)}(Q).
}

\noindent This kind of simplification is very powerful in recursive calculations. We can then imagine repeating the previous examplitude supersymmetrically. The only other factor shifted is $\hat{x}$, and so the recursion is identical to the scalar case (\emph{c.f.} \cite{Badger_2005, Ochirov_2018}),
\eq{
A[\Phi,2^+,3^+,\Phi] = \delta^{(2)}(Q)\frac{m   [\tilde{2}\tilde{3}]}{\langle23\rangle(p_{12}^2-m^2)}.
}
Supersymmetric bookkeeping gives the scalar and fermionic amplitudes in one shot. At higher points the fermionic deltas pose no obstruction as long as we always shift the massive legs, and so computing the \emph{superamplitude} is again identical to the scalar recursion, except for the spectator fermionic deltas. We then recover the known result (see \cite{Ochirov_2018} where it was derived instead from \emph{massless} BCFW-recursion),
\eq{
A[\Phi, 2^+,\cdots,n\!-\!1^+,\Phi] = \frac{m\delta^{(2)}(Q) [\tilde{2}|\Pi_{j=2}^{n-3}\big(p_{12\cdots j}p_{j+1}+(p_{12\cdots j}^2-m^2)\big)|\tilde{n\!-\!1}]}{\Pi_{j=2}^{n-2}\langle j|j\!+\!1\rangle(p_{12\cdots j}^2-m^2)}.
}
\noindent Of course, the $(n-2)$-minus amplitude can be derived by the same manipulations. 
\eq{
A[\Phi, 2^-,\cdots,n\!-\!1^-,\Phi] = \frac{m\delta^{(2)}(\tilde{Q}) \langle 2 |\Pi_{j=2}^{n-3}\big(p_{12\cdots j}p_{j+1}+(p_{12\cdots j}^2-m^2)\big)|n\!-\!1\rangle}{\Pi_{j=2}^{n-2}[\tilde{j}|\tilde{j\!+\!1}](p_{12\cdots j}^2-m^2)}.
}
This provides a nice retrospective on why people were able to construct $2$-massive $(n-2)$-plus amplitudes by simply `covariantizing' the scalar expression \cite{Forde_2006, Ferrario_2006, Schwinn_2007, Ochirov_2018, Lazopoulos_2022}. The amplitudes discussed in those papers involve adjacent massive legs, but the supersymmetric argument works also for non-adjacent massive legs just as well.\vspace{\baselineskip}

\section{Conclusion and Outlook}
\label{sec:conclusion}

In this work we have introduced a new formula for the BCFW-shift using the covariant massive spinor-helicity formalism of \cite{Arkani-Hamed:2017jhn} applicable to any pair of masses and with spinors prepared in any basis, which unifies all known BCFW-shifts in the literature as special cases of our general expression. We have also introduced a new construction that trivializes the preparation of massive spinors given a momentum, and allows one to adopt convenient bases in which recursion may be simplified. Besides providing a more general expression for the massive shift than was previously known (and making such a recursive computation easier to implement), these two results immediately suggest a simplified version of the mixed-mass BCFW-shift. In appropriate bases the spinors for \emph{all} combinations of masses are shifted identically. However, the preparation of the ``appropriate basis'' for the massive case may be sufficiently onerous to make the general form of the shift more useful in computations. 

The obvious next step would be to apply this method to the calculation of amplitudes in specific theories of interest, such as QED, but the lack of colour-ordering makes the viability of on-shell recursion suspect. Recent works (see \cite{Melia:2013bta, Johansson:2015oia, bourjaily2026treeamplitudeschargedmatter}) have explicitly illustrated how massless QED tree amplitudes may be obtained from massless QCD tree amplitudes, and that QED amplitudes surprisingly \emph{can} be thought of as having a notion of colour-ordering. A similar statement holds for massive QED as well, so having good computational control of the partial amplitudes of QCD is enough to reconstruct QED amplitudes. Amplitudes of gravitons and massive fermions can also be obtained from QCD \cite{Johansson:2015oia, Johansson_2019}. 

However, this is far from the last hurdle one must contend with. As discussed in \cite{Melia:2013bta, Kalin:2017oqr, bourjaily2026treeamplitudeschargedmatter}, amplitudes involving matter lines are organized by the combinatorics of Dyck words, which depend strongly on the (in)distinguishability of the matter (the \emph{flavours} of matter involved). Determination of which BCFW-shifts are ``good'' and ``bad'' also depends on the flavours involved. Furthermore, whether or not the masses of distinct flavours of matter are identical or not can make a big difference in the analysis. These are, of course, surmountable complications, but they illustrate that subtleties abound in this program, and care must be taken.

Given the phenomenological interest of the above theories, an efficient package for computing their amplitudes would be valuable indeed. Due to the theory-dependent analysis necessary---whereas everything we have presented herein applies quite broadly---we leave this to future works. Nevertheless, preliminary applications of this formalism have allowed us to compute in short order ten-particle partial amplitudes involving massive-scalars minimally coupled to gluons with \emph{arbitrary} helicity configurations. Such amplitudes with limited helicity configurations (\emph{e.g.} all-plus) have been studied using massless shifts (see \cite{Forde_2006}), but a more general and computationally friendly paradigm for recursion involving massive particles will allow us to consider wider classes of amplitudes (\emph{e.g.} amplitudes involving \emph{no} massless legs).

Lastly, we remark that spinor-helicity formalisms have been constructed in $D>4$ \cite{Chiodaroli_2023, pokraka2024fivedimensionalspinorhelicitymasses, Cheung:2009dc, Caron-Huot_2011_10d}, and non-covariant BCFW in higher dimensions was discussed all the way back in \cite{ArkaniHamed:2008yf}. In $D>4$ we can argue geometrically that the BCFW-shift is no longer unique: since $(p_a\wedge p_b)^{\perp}$ is a $D-2$ dimensional space, after imposing nullity of the shift vector $r$ we find a $D-3$ dimensional space. Since we shift by $zr$ the definition of $z$ fixes the overall scale, and $D-4$ gives a unique solution in four dimensions, but suggests that one has some freedom in choosing the shift in higher dimensions. It would then be interesting to explore a covariant formulation of the shift in higher dimensions. The obstacle to this is that spinor-helicity formalisms are dimension-specific, which makes uniform treatment difficult. A uniform construction would then be useful, even if it does not exploit all of the available dimension-specific specialties. This would also simplify the implementation of spinor-helicity in higher dimensions tremendously, and we leave this to future works.

\vfill
\section*{Acknowledgments}%
\vspace{0pt}
\noindent The authors gratefully acknowledge helpful conversations with Nima Arkani-Hamed, Radu Roiban, Philip Velie, and Cristian Vergu, and the hospitality of the Institute for Advanced Study, Princeton. This work was supported in part by a grant from the US Department of Energy (No.\ DE-SC00019066) and the J.~Robert Oppenheimer Visiting Professor Endowment at the Institute for Advanced Study, Princeton. 
%
%

\begin{thebibliography}{10}

\bibitem{Giele:1993dj}
W.~T. Giele, E.~W.~N. Glover, and D.~A. Kosower, ``{Higher Order Corrections to
  Jet Cross-Sections in Hadron Colliders},''
  \href{http://dx.doi.org/10.1016/0550-3213(93)90365-V}{{\em Nucl. Phys. B}
  {\bf 403} (1993)  633--670}, \href{http://arxiv.org/abs/hep-ph/9302225}{{
  arXiv:hep-ph/9302225}}.

\bibitem{Mangano:1990by}
M.~L. Mangano and S.~J. Parke, ``{Multiparton Amplitudes in Gauge Theories},''
  \href{http://dx.doi.org/10.1016/0370-1573(91)90091-Y}{{\em Phys. Rept.} {\bf
  200} (1991)  301--367}, \href{http://arxiv.org/abs/hep-th/0509223}{{
  arXiv:hep-th/0509223}}.

\bibitem{Bern:2007dw}
Z.~Bern, L.~J. Dixon, and D.~A. Kosower, ``{On-Shell Methods in Perturbative
  QCD},'' \href{http://dx.doi.org/10.1016/j.aop.2007.04.014}{{\em Annals Phys.}
  {\bf 322} (2007)  1587--1634},
\href{http://arxiv.org/abs/0704.2798}{{ arXiv:0704.2798 [hep-ph]}}.

\bibitem{buonanno2022snowmasswhitepapergravitational}
A.~Buonanno, M.~Khalil, D.~O'Connell, R.~Roiban, M.~P. Solon, and M.~Zeng,
  ``{Snowmass White Paper: Gravitational Waves and Scattering Amplitudes},''
  2022.
\newblock \url{https://arxiv.org/abs/2204.05194}.

\bibitem{Bern:2019crd}
Z.~Bern, C.~Cheung, R.~Roiban, C.-H. Shen, M.~P. Solon, and M.~Zeng, ``{Black
  Hole Binary Dynamics from the Double Copy and Effective Theory},''
  \href{http://dx.doi.org/10.1007/JHEP10(2019)206}{{\em JHEP} {\bf 10} (2019)
  206}, \href{http://arxiv.org/abs/1908.01493}{{ arXiv:1908.01493 [hep-th]}}.

\bibitem{Adamo_2024}
T.~Adamo, R.~Gonzo, and A.~Ilderton, ``{Gravitational Bound Waveforms from
  Amplitudes},'' \href{http://dx.doi.org/10.1007/JHEP05(2024)034}{{\em JHEP}
  {\bf 05} (2024)  034}, \href{http://arxiv.org/abs/2402.00124}{{
  arXiv:2402.00124 [hep-th]}}.

\bibitem{vanderWaerden:1929}
B.~L. van~der Waerden, ``{Spinoranalyse},'' {\em Nach. Ges. Wiss. G\"{o}ttingen
  Math.-Phys.} {\bf 1} (1929)  100--109.

\bibitem{Kleiss:1985yh}
R.~Kleiss and W.~J. Stirling, ``{Spinor Techniques for Calculating
  $p\,\bar{p}\to W^\pm Z^0\,+\mathrm{Jets}$},''
\href{http://dx.doi.org/10.1016/0550-3213(85)90285-8}{{\em Nucl. Phys.} {\bf
  B262} (1985)  235--262}.

\bibitem{Parke:1986gb}
S.~J. Parke and T.~R. Taylor, ``{An Amplitude for $n$-Gluon Scattering},''
\href{http://dx.doi.org/10.1103/PhysRevLett.56.2459}{{\em Phys. Rev. Lett.}
  {\bf 56} (1986)  2459}.

\bibitem{Berends:1987me}
F.~A. Berends and W.~T. Giele, ``{Recursive Calculations for Processes with $n$
  Gluons},''
\href{http://dx.doi.org/10.1016/0550-3213(88)90442-7}{{\em Nucl. Phys.} {\bf
  B306} (1988)  759}.

\bibitem{Duca1999NewCD}
V.~del Duca, L.~J. Dixon, and F.~Maltoni, ``{New Color Decompositions for Gauge
  Amplitudes at Tree and Loop Level},'' {\em Nuclear Physics} {\bf 571} (1999)
  51--70. \url{https://api.semanticscholar.org/CorpusID:1602861}.

\bibitem{Witten:2003nn}
E.~Witten, ``{Perturbative Gauge Theory as a String Theory in Twistor Space},''
  \href{http://dx.doi.org/10.1007/s00220-004-1187-3}{{\em Commun. Math. Phys.}
  {\bf 252} (2004)  189--258},
\href{http://arxiv.org/abs/hep-th/0312171}{{ arXiv:hep-th/0312171}}.

\bibitem{BCF}
R.~Britto, F.~Cachazo, and B.~Feng, ``{New Recursion Relations for Tree
  Amplitudes of Gluons},''
  \href{http://dx.doi.org/10.1016/j.nuclphysb.2005.02.030}{{\em Nucl.Phys.}
  {\bf B715} (2005)  499--522},
\href{http://arxiv.org/abs/hep-th/0412308}{{ arXiv:hep-th/0412308 [hep-th]}}.

\bibitem{BCFW}
R.~Britto, F.~Cachazo, B.~Feng, and E.~Witten, ``{Direct Proof of Tree-Level
  Recursion Relation in Yang- Mills Theory},''
  \href{http://dx.doi.org/10.1103/PhysRevLett.94.181602}{{\em Phys. Rev. Lett.}
  {\bf 94} (2005)  181602},
\href{http://arxiv.org/abs/hep-th/0501052}{{ arXiv:hep-th/0501052}}.

\bibitem{Bourjaily:2023uln}
J.~L. Bourjaily, ``{Computational Tools for Trees in Gauge Theory and
  Gravity},'' \href{http://arxiv.org/abs/2312.17745}{{ arXiv:2312.17745
  [hep-th]}}.

\bibitem{Bern:2011qt}
Z.~Bern and Y.-t. Huang, ``{Basics of Generalized Unitarity},''
  \href{http://dx.doi.org/10.1088/1751-8113/44/45/454003}{{\em J. Phys.} {\bf
  A44} (2011)  454003},
\href{http://arxiv.org/abs/1103.1869}{{ arXiv:1103.1869 [hep-th]}}.

\bibitem{Bourjaily:2017wjl}
J.~L. Bourjaily, E.~Herrmann, and J.~Trnka, ``{Prescriptive Unitarity},''
  \href{http://dx.doi.org/10.1007/JHEP06(2017)059}{{\em JHEP} {\bf 06} (2017)
  059},
\href{http://arxiv.org/abs/1704.05460}{{ arXiv:1704.05460 [hep-th]}}.

\bibitem{PositiveGrassmannianBook}
N.~Arkani-Hamed, J.~L. Bourjaily, F.~Cachazo, A.~B. Goncharov, A.~Postnikov,
  and J.~Trnka, \href{http://dx.doi.org/10.1017/CBO9781316091548}{{\em
  {Grassmannian Geometry of Scattering Amplitudes}}}.
\newblock Cambridge University Press, 2016.

\bibitem{ArkaniHamed:2010gh}
N.~Arkani-Hamed, J.~L. Bourjaily, F.~Cachazo, and J.~Trnka, ``{Local Integrals
  for Planar Scattering Amplitudes},''
  \href{http://dx.doi.org/10.1007/JHEP06(2012)125}{{\em JHEP} {\bf 1206} (2012)
   125},
\href{http://arxiv.org/abs/1012.6032}{{ arXiv:1012.6032 [hep-th]}}.

\bibitem{ArkaniHamed:2010kv}
N.~Arkani-Hamed, J.~L. Bourjaily, F.~Cachazo, S.~Caron-Huot, and J.~Trnka,
  ``{The All-Loop Integrand For Scattering Amplitudes in Planar
  $\mathcal{N}\!=\!4$ sYM},''
  \href{http://dx.doi.org/10.1007/JHEP01(2011)041}{{\em JHEP} {\bf 1101} (2011)
   041},
\href{http://arxiv.org/abs/1008.2958}{{ arXiv:1008.2958 [hep-th]}}.

\bibitem{Arkani-Hamed:2014dca}
N.~Arkani-Hamed, A.~Hodges, and J.~Trnka, ``{Positive Amplitudes In The
  Amplituhedron},'' \href{http://dx.doi.org/10.1007/JHEP08(2015)030}{{\em JHEP}
  {\bf 08} (2015)  030},
\href{http://arxiv.org/abs/1412.8478}{{ arXiv:1412.8478 [hep-th]}}.

\bibitem{Arkani-Hamed:2013jha}
N.~Arkani-Hamed and J.~Trnka, ``{The Amplituhedron},''
  \href{http://dx.doi.org/10.1007/JHEP10(2014)030}{{\em JHEP} {\bf 1410} (2014)
   30},
\href{http://arxiv.org/abs/1312.2007}{{ arXiv:1312.2007 [hep-th]}}.

\bibitem{BCJ}
Z.~Bern, J.~Carrasco, and H.~Johansson, ``{New Relations for Gauge-Theory
  Amplitudes},'' \href{http://dx.doi.org/10.1103/PhysRevD.78.085011}{{\em Phys.
  Rev.} {\bf D78} (2008)  085011},
\href{http://arxiv.org/abs/0805.3993}{{ arXiv:0805.3993 [hep-ph]}}.

\bibitem{Bern:2010ue}
Z.~Bern, J.~J.~M. Carrasco, and H.~Johansson, ``{Perturbative Quantum Gravity
  as a Double Copy of Gauge Theory},''
  \href{http://dx.doi.org/10.1103/PhysRevLett.105.061602}{{\em Phys. Rev.
  Lett.} {\bf 105} (2010)  061602},
\href{http://arxiv.org/abs/1004.0476}{{ arXiv:1004.0476 [hep-th]}}.

\bibitem{Bourjaily:2023ycy}
J.~L. Bourjaily, N.~Kalyanapuram, K.~Patatoukos, M.~Plesser, and Y.~Zhang,
  ``{Gauge-Invariant Double Copies via Recursion},''
  \href{http://dx.doi.org/10.1103/PhysRevLett.131.191601}{{\em Phys. Rev.
  Lett.} {\bf 131} (2023) no. 19, 191601},
  \href{http://arxiv.org/abs/2307.02542}{{ arXiv:2307.02542 [hep-th]}}.

\bibitem{Bern:1995db}
Z.~Bern and A.~G. Morgan, ``{Massive Loop Amplitudes from Unitarity},''
  \href{http://dx.doi.org/10.1016/0550-3213(96)00078-8}{{\em Nucl. Phys.} {\bf
  B467} (1996)  479--509},
\href{http://arxiv.org/abs/hep-ph/9511336}{{ arXiv:hep-ph/9511336 [hep-ph]}}.

\bibitem{Bern:2010qa}
Z.~Bern, J.~J. Carrasco, T.~Dennen, Y.-t. Huang, and H.~Ita, ``{Generalized
  Unitarity and Six-Dimensional Helicity},''
  \href{http://dx.doi.org/10.1103/PhysRevD.83.085022}{{\em Phys. Rev.} {\bf
  D83} (2011)  085022},
\href{http://arxiv.org/abs/1010.0494}{{ arXiv:1010.0494 [hep-th]}}.

\bibitem{Elvang:2011ub}
H.~Elvang, D.~Z. Freedman, and M.~Kiermaier, ``{Integrands for QCD Rational
  Terms and $\mathcal{N}\!=\!4$ sYM from Massive CSW Rules},''
  \href{http://dx.doi.org/10.1007/JHEP06(2012)015}{{\em JHEP} {\bf 1206} (2012)
   015},
\href{http://arxiv.org/abs/1111.0635}{{ arXiv:1111.0635 [hep-th]}}.

\bibitem{Arkani-Hamed:2017jhn}
N.~Arkani-Hamed, T.-C. Huang, and Y.-t. Huang, ``{Scattering Amplitudes for All
  Masses and Spins},'' \href{http://dx.doi.org/10.1007/JHEP11(2021)070}{{\em
  JHEP} {\bf 11} (2021)  070}, \href{http://arxiv.org/abs/1709.04891}{{
  arXiv:1709.04891 [hep-th]}}.

\bibitem{Badger_2005}
S.~D. Badger, E.~W.~N. Glover, V.~V. Khoze, and P.~Svr\v{c}ek, ``{Recursion
  Relations for Gauge Theory Amplitudes with Massive Particles},''
  \href{http://dx.doi.org/10.1088/1126-6708/2005/07/025}{{\em JHEP} {\bf 07}
  (2005)  025}, \href{http://arxiv.org/abs/hep-th/0504159}{{
  arXiv:hep-th/0504159}}.

\bibitem{Schwinn_2007}
C.~Schwinn and S.~Weinzierl, ``{On-Shell Recursion Relations for All Born QCD
  Amplitudes},'' \href{http://dx.doi.org/10.1088/1126-6708/2007/04/072}{{\em
  JHEP} {\bf 04} (2007)  072}, \href{http://arxiv.org/abs/hep-ph/0703021}{{
  arXiv:hep-ph/0703021}}.

\bibitem{Ballav_2021}
S.~Ballav and A.~Manna, ``{Recursion Relations for Scattering Amplitudes with
  Massive Particles},'' \href{http://dx.doi.org/10.1007/JHEP03(2021)295}{{\em
  JHEP} {\bf 03} (2021)  295}, \href{http://arxiv.org/abs/2010.14139}{{
  arXiv:2010.14139 [hep-th]}}.

\bibitem{Wu_2022}
C.~Wu and S.-H. Zhu, ``{Massive On-Shell Recursion Relations for $n$-Point
  Amplitudes},'' \href{http://dx.doi.org/10.1007/JHEP06(2022)117}{{\em JHEP}
  {\bf 06} (2022)  117}, \href{http://arxiv.org/abs/2112.12312}{{
  arXiv:2112.12312 [hep-th]}}.

\bibitem{Elvang:2013cua}
H.~Elvang and Y.-t. Huang, ``{Scattering Amplitudes},''
\href{http://arxiv.org/abs/1308.1697}{{ arXiv:1308.1697 [hep-th]}}.

\bibitem{Weinberg:1995mt}
S.~Weinberg, {\em {The Quantum Theory of Fields. Vol. 1: Foundations}}.
\newblock Cambridge University Press,
1995.
\newblock

\bibitem{Moynihan_2018}
N.~Moynihan and J.~Murugan, ``{Comments on Scattering in Massive Gravity, vDVZ
  and BCFW},'' \href{http://dx.doi.org/10.1088/1361-6382/aacb73}{{\em Class.
  Quant. Grav.} {\bf 35} (2018) no. 15, 155005},
  \href{http://arxiv.org/abs/1711.03956}{{ arXiv:1711.03956 [hep-th]}}.

\bibitem{burger2019amplitudesgravityclassicaldiscontinuities}
D.~J. Burger, R.~Carballo-Rubio, N.~Moynihan, J.~Murugan, and A.~Weltman,
  ``{Amplitudes, Gravity and Classical Discontinuities},'' 2019.
\newblock \url{https://arxiv.org/abs/1905.05128}.

\bibitem{burger2020onshellperspectivesmasslesslimit}
D.~J. Burger, N.~Moynihan, and J.~Murugan, ``{On-Shell Perspectives on the
  Massless Limit of Massive Supergravity},'' 2020.
\newblock \url{https://arxiv.org/abs/2005.14077}.

\bibitem{Pittau_1998}
R.~Pittau, ``{A Simple Method for Multi-Leg Loop Calculations 2: A General
  Algorithm},''
  \href{http://dx.doi.org/https://doi.org/10.1016/S0010-4655(98)00040-X}{{\em
  Computer Physics Communications} {\bf 111} (1998) no. 1, 48--52}.
  \url{https://www.sciencedirect.com/science/article/pii/S001046559800040X}.

\bibitem{Aguila_2004}
F.~del Aguila and R.~Pittau, ``{Recursive Numerical Calculus of One-Loop Tensor
  Integrals},'' \href{http://dx.doi.org/10.1088/1126-6708/2004/07/017}{{\em
  JHEP} {\bf 07} (2004)  017}, \href{http://arxiv.org/abs/hep-ph/0404120}{{
  arXiv:hep-ph/0404120}}.

\bibitem{OPP}
G.~Ossola, C.~G. Papadopoulos, and R.~Pittau, ``{Reducing Full One-Loop
  Amplitudes to Scalar Integrals at the Integrand Level},''
  \href{http://dx.doi.org/10.1016/j.nuclphysb.2006.11.012}{{\em Nucl. Phys.}
  {\bf B763} (2007)  147--169},
\href{http://arxiv.org/abs/hep-ph/0609007}{{ arXiv:hep-ph/0609007}}.

\bibitem{ArkaniHamed:2008yf}
N.~Arkani-Hamed and J.~Kaplan, ``{On Tree Amplitudes in Gauge Theory and
  Gravity},'' \href{http://dx.doi.org/10.1088/1126-6708/2008/04/076}{{\em JHEP}
  {\bf 0804} (2008)  076},
\href{http://arxiv.org/abs/0801.2385}{{ arXiv:0801.2385 [hep-th]}}.

\bibitem{Cohen:2010mi}
T.~Cohen, H.~Elvang, and M.~Kiermaier, ``{On-Shell Constructibility of Tree
  Amplitudes in General Field Theories},''
  \href{http://dx.doi.org/10.1007/JHEP04(2011)053}{{\em JHEP} {\bf 1104} (2011)
   053},
\href{http://arxiv.org/abs/1010.0257}{{ arXiv:1010.0257 [hep-th]}}.

\bibitem{Damour_2016}
T.~Damour, ``{Gravitational Scattering, Post-Minkowskian Approximation and
  Effective One-Body Theory},''
  \href{http://dx.doi.org/10.1103/PhysRevD.94.104015}{{\em Phys. Rev. D} {\bf
  94} (2016) no. 10, 104015}, \href{http://arxiv.org/abs/1609.00354}{{
  arXiv:1609.00354 [gr-qc]}}.

\bibitem{Drummond:2008cr}
J.~M. Drummond and J.~M. Henn, ``{All Tree-Level Amplitudes in
  $\mathcal{N}\!=\!4$ SYM},''
  \href{http://dx.doi.org/10.1088/1126-6708/2009/04/018}{{\em JHEP} {\bf 04}
  (2009)  018},
\href{http://arxiv.org/abs/0808.2475}{{ arXiv:0808.2475 [hep-th]}}.

\bibitem{Dixon:2010ik}
L.~J. Dixon, J.~M. Henn, J.~Plefka, and T.~Schuster, ``{All Tree-Level
  Amplitudes in Massless QCD},''
  \href{http://dx.doi.org/10.1007/JHEP01(2011)035}{{\em JHEP} {\bf 1101} (2011)
   035},
\href{http://arxiv.org/abs/1010.3991}{{ arXiv:1010.3991 [hep-ph]}}.

\bibitem{Bourjaily:2010wh}
J.~L. Bourjaily, ``{Efficient Tree-Amplitudes in $\mathcal{N}\!=\!4$: Automatic
  BCFW Recursion in {\sc Mathematica}},''
\href{http://arxiv.org/abs/1011.2447}{{ arXiv:1011.2447 [hep-ph]}}.

\bibitem{Herderschee_2019_mos}
A.~Herderschee, S.~Koren, and T.~Trott, ``{Massive On-Shell Supersymmetric
  Scattering Amplitudes},''
  \href{http://dx.doi.org/10.1007/JHEP10(2019)092}{{\em JHEP} {\bf 10} (2019)
  092}, \href{http://arxiv.org/abs/1902.07204}{{ arXiv:1902.07204 [hep-th]}}.

\bibitem{Herderschee_2019_cba}
A.~Herderschee, S.~Koren, and T.~Trott, ``{Constructing $ \mathcal{N}{=}4$
  Coulomb Branch Superamplitudes},''
  \href{http://dx.doi.org/10.1007/JHEP08(2019)107}{{\em JHEP} {\bf 08} (2019)
  107}, \href{http://arxiv.org/abs/1902.07205}{{ arXiv:1902.07205 [hep-th]}}.

\bibitem{Abhishek_2022}
M.~Abhishek, S.~Hegde, D.~P. Jatkar, and A.~P. Saha, ``{Scattering Amplitudes
  and BCFW in $\mathcal{N}{=}2^*$ Theory},''
  \href{http://dx.doi.org/10.21468/SciPostPhys.13.1.008}{{\em SciPost Phys.}
  {\bf 13} (2022) no. 1, 008}, \href{http://arxiv.org/abs/2202.12204}{{
  arXiv:2202.12204 [hep-th]}}.

\bibitem{Ochirov_2018}
A.~Ochirov, ``{Helicity Amplitudes for QCD with Massive Quarks},''
  \href{http://dx.doi.org/10.1007/JHEP04(2018)089}{{\em JHEP} {\bf 04} (2018)
  089}, \href{http://arxiv.org/abs/1802.06730}{{ arXiv:1802.06730 [hep-ph]}}.

\bibitem{Forde_2006}
D.~Forde and D.~A. Kosower, ``{All-Multiplicity Amplitudes with Massive
  Scalars},'' \href{http://dx.doi.org/10.1103/PhysRevD.73.065007}{{\em Phys.
  Rev. D} {\bf 73} (2006)  065007},
  \href{http://arxiv.org/abs/hep-th/0507292}{{ arXiv:hep-th/0507292}}.

\bibitem{Ferrario_2006}
P.~Ferrario, G.~Rodrigo, and P.~Talavera, ``{Compact Multigluonic Scattering
  Amplitudes with Heavy Scalars and Fermions},''
  \href{http://dx.doi.org/10.1103/PhysRevLett.96.182001}{{\em Phys. Rev. Lett.}
  {\bf 96} (2006)  182001}, \href{http://arxiv.org/abs/hep-th/0602043}{{
  arXiv:hep-th/0602043}}.

\bibitem{Lazopoulos_2022}
A.~Lazopoulos, A.~Ochirov, and C.~Shi, ``{All-Multiplicity Amplitudes with Four
  Massive Quarks and Identical-Helicity Gluons},''
  \href{http://dx.doi.org/10.1007/JHEP03(2022)009}{{\em JHEP} {\bf 03} (2022)
  009}, \href{http://arxiv.org/abs/2111.06847}{{ arXiv:2111.06847 [hep-th]}}.

\bibitem{Melia:2013bta}
T.~Melia, ``{Dyck Words and Multiquark Primitive Amplitudes},''
  \href{http://dx.doi.org/10.1103/PhysRevD.88.014020}{{\em Phys. Rev. D} {\bf
  88} (2013) no. 1, 014020}, \href{http://arxiv.org/abs/1304.7809}{{
  arXiv:1304.7809 [hep-ph]}}.

\bibitem{Johansson:2015oia}
H.~Johansson and A.~Ochirov, ``{Color-Kinematics Duality for QCD Amplitudes},''
  \href{http://dx.doi.org/10.1007/JHEP01(2016)170}{{\em JHEP} {\bf 01} (2016)
  170}, \href{http://arxiv.org/abs/1507.00332}{{ arXiv:1507.00332 [hep-ph]}}.

\bibitem{bourjaily2026treeamplitudeschargedmatter}
J.~L. Bourjaily, M.~Plesser, and P.~Velie, ``{Tree Amplitudes with Charged
  Matter in Pure Gauge Theory},'' 2026.
\newblock \url{https://arxiv.org/abs/2604.18672}.

\bibitem{Johansson_2019}
H.~Johansson and A.~Ochirov, ``{Double Copy for Massive Quantum Particles with
  Spin},'' \href{http://dx.doi.org/10.1007/JHEP09(2019)040}{{\em JHEP} {\bf 09}
  (2019)  040}, \href{http://arxiv.org/abs/1906.12292}{{ arXiv:1906.12292
  [hep-th]}}.

\bibitem{Kalin:2017oqr}
G.~K{\"a}lin, ``{Cyclic Mario Worlds---Color-Decomposition for One-Loop QCD},''
  \href{http://dx.doi.org/10.1007/JHEP04(2018)141}{{\em JHEP} {\bf 04} (2018)
  141}, \href{http://arxiv.org/abs/1712.03539}{{ arXiv:1712.03539 [hep-th]}}.

\bibitem{Chiodaroli_2023}
M.~Chiodaroli, M.~G\"{u}naydin, H.~Johansson, and R.~Roiban, ``{Spinor-Helicity
  Formalism for Massive and Massless Amplitudes in Five Dimensions},''
  \href{http://dx.doi.org/10.1007/JHEP02(2023)040}{{\em JHEP} {\bf 02} (2023)
  040}, \href{http://arxiv.org/abs/2202.08257}{{ arXiv:2202.08257 [hep-th]}}.

\bibitem{pokraka2024fivedimensionalspinorhelicitymasses}
A.~Pokraka, S.~Rajan, L.~Ren, A.~Volovich, and W.~W. Zhao, ``{Five-Dimensional
  Spinor Helicity for All Masses and Spins},'' 2024.
\newblock \url{https://arxiv.org/abs/2405.09533}.

\bibitem{Cheung:2009dc}
C.~Cheung and D.~O'Connell, ``{Amplitudes and Spinor-Helicity in Six
  Dimensions},'' \href{http://dx.doi.org/10.1088/1126-6708/2009/07/075}{{\em
  JHEP} {\bf 0907} (2009)  075},
\href{http://arxiv.org/abs/0902.0981}{{ arXiv:0902.0981 [hep-th]}}.

\bibitem{Caron-Huot_2011_10d}
S.~Caron-Huot and D.~O'Connell, ``{Spinor Helicity and Dual Conformal Symmetry
  in Ten Dimensions},'' \href{http://dx.doi.org/10.1007/JHEP08(2011)014}{{\em
  JHEP} {\bf 08} (2011)  014}, \href{http://arxiv.org/abs/1010.5487}{{
  arXiv:1010.5487 [hep-th]}}.

\end{thebibliography}

\addcontentsline{toc}{section}{References}
\providecommand{\href}[2]{#2}\begingroup\raggedright\endgroup
\end{document}